\documentclass[runningheads]{llncs}
\usepackage{graphicx}
\usepackage[colorlinks=true,
            urlcolor=blue,
            citecolor=blue,
            linkcolor=blue
            ]{hyperref}
\usepackage{caption}
\usepackage{afterpage}

\begin{document}
%

\title{\ortoyw: Modeling and Visualizing OpenRefine Histories as YesWorkflow Diagrams}
\titlerunning{\ortoyw: From OpenRefine Histories to YesWorkflow Diagrams}

\author{Nikolaus Nova Parulian\inst{1} \and Lan Li\inst{1} \and  Bertram Lud\"ascher\inst{1}}
\authorrunning{Nikolaus Parulian \and Lan Li \and Bertram Lud\"ascher}

\institute{
Center for Informatics Research in Science \& Scholarship \\
School of Information Sciences\\University of Illinois at Urbana-Champaign, USA\\
\email{\{nnp2,lanl2,ludaesch\}@illinois.edu}
}

\newcommand{\ortoyw}{\textsf{or2yw}}


%
\maketitle              
\begin{abstract}

  OpenRefine is a popular open-source data cleaning tool. It allows users to export a previously executed data cleaning workflow in
  a JSON format for possible reuse on other datasets. We have developed \ortoyw, a novel tool that maps a JSON-formatted OpenRefine {operation history} to a YesWorkflow (YW) model, which then can be visualized and queried using the YW tool.
  The latter was originally developed to allow researchers a simple way to annotate their program scripts in order to reveal the
  workflow steps and dataflow dependencies implicit in those scripts. With \ortoyw\ the user can {automatically} generate YW
  models from OpenRefine operation histories, thus providing a ``workflow view'' on a previously executed sequence of data cleaning operations.

  The \ortoyw\ tool can generate different types of YesWorkflow models, e.g., a \emph{linear model} which mirrors the sequential
  execution order of operations in OpenRefine, and a \emph{parallel model} which reveals independent workflow branches, based on a
  simple analysis of dependencies between steps: if two operations are independent of each other (e.g., when the columns they read and write do not overlap) then these can be viewed as parallel steps in the data cleaning
  workflow.
  The resulting YW models can be understood as a form of \emph{prospective provenance}, i.e., knowledge artifacts that can be
  queried and visualized (i) to help authors document their own data cleaning workflows, thereby increasing transparency, and (ii)~to help other
  users, who might want to reuse such workflows, to understand them better.
\end{abstract}

\section{Introduction: OpenRefine Histories as Provenance}

\begin{figure}[t]
\centering
\fbox{\includegraphics[width=0.85\textwidth]{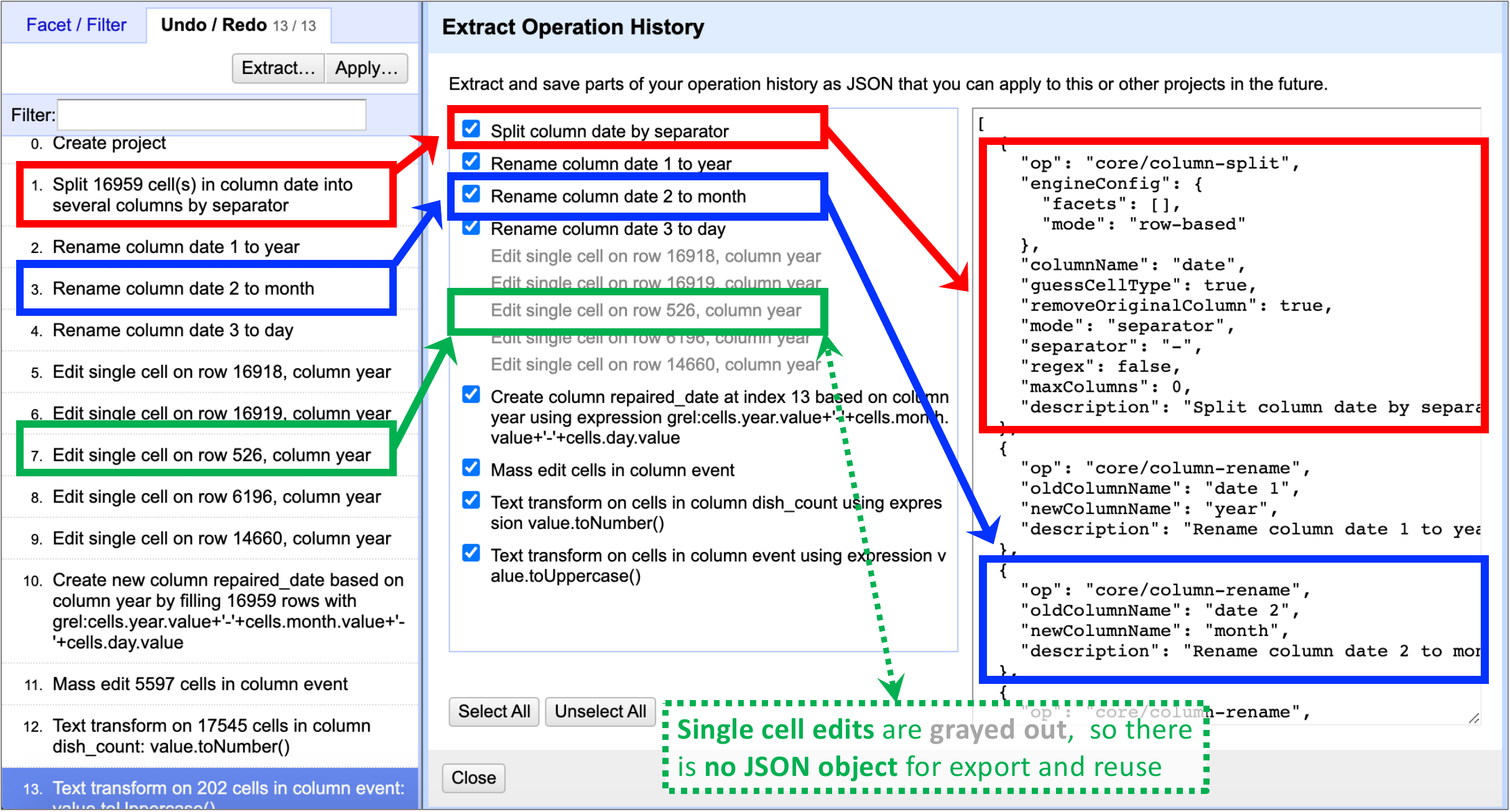}}
\caption{\textbf{OpenRefine operation history} ({left}); checklist ({center}); and JSON objects for export and reuse ({right}). After clicking \texttt{Extract} near the upper left, a user can select the operations to be exported with a checklist. The selected operations appear in the JSON panel on the right. Unlike other operations, \textsf{single cell} edits are {not} user-selectable and thus not included in the JSON-formatted history to be exported: while other operations apply to all cells in a column uniformly (and hence are reusable), a single cell edit is by design a ``one-off'' update that is not reusable in another data cleaning recipe.}
\label{fig:recipe}
\end{figure}

OpenRefine is an open-source data cleaning tool that allows users to execute data transformations in a browser-based, spreadsheet-like GUI \cite{openrefine2020}. 
OpenRefine also records the data cleaning operations used in a project in an \emph{operation history} (Fig.\,\ref{fig:recipe}). This history can be understood as a form of \emph{provenance} because it reveals what operations were executed on the dataset.

Provenance information has many uses and applications \cite {herschel2017survey}, e.g., to increase transparency (and thus also improve the reproducibility) of workflows, to aid quality estimation, and more generally to contribute to the trustworthiness of workflow products.  We can distinguish different types of provenance information:

\begin{itemize}
\item \textbf{Prospective provenance} describes the general \emph{workflow} (or ``recipe'') that has been (or is being) used for cleaning the data, without reference to details that may vary with each workflow run. 
\item \textbf{Retrospective provenance}, in contrast, describes specific details of workflow runs that \emph{have been} executed and that are \emph{not} common across all runs.
\end{itemize} 

\noindent The separation between both forms of provenance is not necessarily strict. While OpenRefine histories can be understood as carrying workflow information (i.e., \emph{prospective} provenance), there are also elements of \emph{retrospective} provenance buried in various OpenRefine artifacts, e.g., in history and internal project files. Using both kinds of provenance together in a \emph{hybrid} provenance model increases the expressive power and utility of both forms of provenance \cite{mcphillips2015retrospective,pimentel2016yinyang,zhang2017revealing,zhang2017using}.

As part of its operation history, OpenRefine records names of data transformations, the linear sequence in which operations were executed, and a description of each execution step. This information can also be understood quite naturally as prospective provenance. On the other hand, the number of rows or cells affected by an operation constitutes retrospective provenance.

Fig.\,\ref{fig:recipe} depicts pieces of provenance information from the operation history checklist. The operation described in the highlighted red box explains the first step from the transformation list: it executed a \textsf{column-split} operation on column \textsf{date}, resulting in 16,959 cell changes.
Additional provenance captured by OpenRefine includes: which columns were used to produce which other columns, and the specific function parameters and arguments\footnote {Similar details are captured by \textsf{noWorkflow} \cite{murta2014noworkflow} for Python function calls.} used during runtime. The latter can be viewed as both prospective and retrospective provenance: In Fig.\,\ref{fig:recipe} on the right, the \textsf{column-rename} operation was applied on column \textsf{date 2}. The operation identified by the blue arrow shows more details of parameters of this execution including the \textsf{oldColumnName} and \textsf{newColumnName} of this transformation.


The extracted OpenRefine history is a powerful feature that allows users to reuse some or all steps of a data cleaning workflow on a different dataset as long as the new dataset has the same (or at least a compatible) data schema. For single cell edits, however, i.e., ``one-off'' operations that have been applied to a specific cell only (rather than the whole column), OpenRefine does not capture provenance in the same way as for reusable operations (green boxes in Fig.\,\ref{fig:recipe}). This export limitation highlights the intended use of the extracted operation history, i.e., as generic ``recipes'' that can be applied uniformly across rows or columns.  A single cell operation only fixes one cell value in the current dataset and thus is not part of the exported JSON recipe.



\subsection*{From Data Cleaning Histories to Workflows}


In the \ortoyw\ approach, we first analyze OpenRefine recipes extracted from histories.  We can distinguish different kinds of transformations, i.e., \emph{intra-column} and \emph{inter-column} operations. The former read and write a single column $C$, independent of other columns; the latter may read or write multiple columns \cite{nunez2020first}.  For example, a \textsf{to-lowercase} transformation applied to column $A$ only depends on the values in $A$ and is thus {independent} of other intra-column operations, e.g., \textsf{to-uppercase} or \textsf{trim-whitespace} performed on a different column $B$.
On the other hand, operations such as \textsf{split-column} or \textsf{merge-columns} affect multiple columns and are not independent of operations that read or write those columns.

\begin{figure}[t]
\centering
\includegraphics[width=1.1\textwidth]{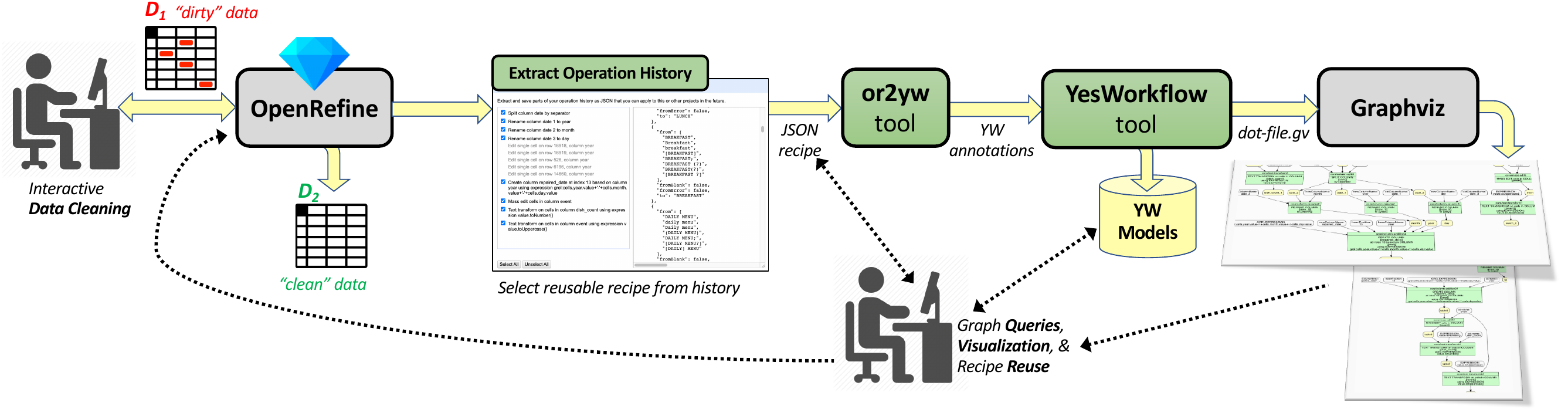} \caption{\textbf{The \ortoyw\ tool in action:} A user cleans dataset \emph{D}$_1$ with OpenRefine, obtaining a clean(er) dataset \emph{D}$_2$. She then extracts from the operation history a JSON recipe, which she feeds into \ortoyw. This generates a YW model that can be queried and visualized (via Graphviz). The original user (a data curator or researcher) and other users can make use of YW models and OpenRefine recipes when developing new data cleaning workflows.}
\label{fig:diagram}
\end{figure} 


For a user, it is not easy to understand the effects of and dependencies between operations when presented in their native JSON form. Tracing provenance information such as column and operation dependencies quickly becomes challenging, especially if the interactive data cleaning workflow is long-running (hours or days) and requires many ``user breaks''.  Motivated by these challenges, we try to address these research questions: 

\begin{itemize} 
\item Can we extract the provenance information provided by an OpenRefine history and generate a corresponding workflow graph?  \item Can we identify and expose independent subworkflows, e.g., by visualizing them as parallel branches in the data cleaning  workflow?   
\end{itemize}

\noindent We answer both questions in the affirmative with our \ortoyw\ tool: the tool allows users to automatically generate YesWorkflow models from OpenRefine histories. These workflow models can then be processed by a YW tool, e.g., to query or visualize the resulting data cleaning workflows \cite{mcphillips2015retrospective,mcphillips2015yesworkflow}.  Thus, the \ortoyw\ tool (Fig.\,\ref{fig:diagram}) helps uncover provenance information by visualizing the operations workflow with its dataflow dependencies and effects on the data schema.

\section{OpenRefine Transformations  and the \ortoyw\ Prototype}
 The \ortoyw\ tool is open source and freely available from Github \cite{or2yw} and PyPI\footnote{\url{https://pypi.org/project/or2ywtool}}. There are currently three different \ortoyw\ workflow options that dictate which type of model is being generated: \emph{linear} (sequential), \emph{parallel}, and \emph{collapsed}.

\subsection*{The Linear Model}

\begin{figure}[th]
  \centering
  \begin{minipage}{.5\linewidth}
    {\includegraphics[height=0.45\textheight]{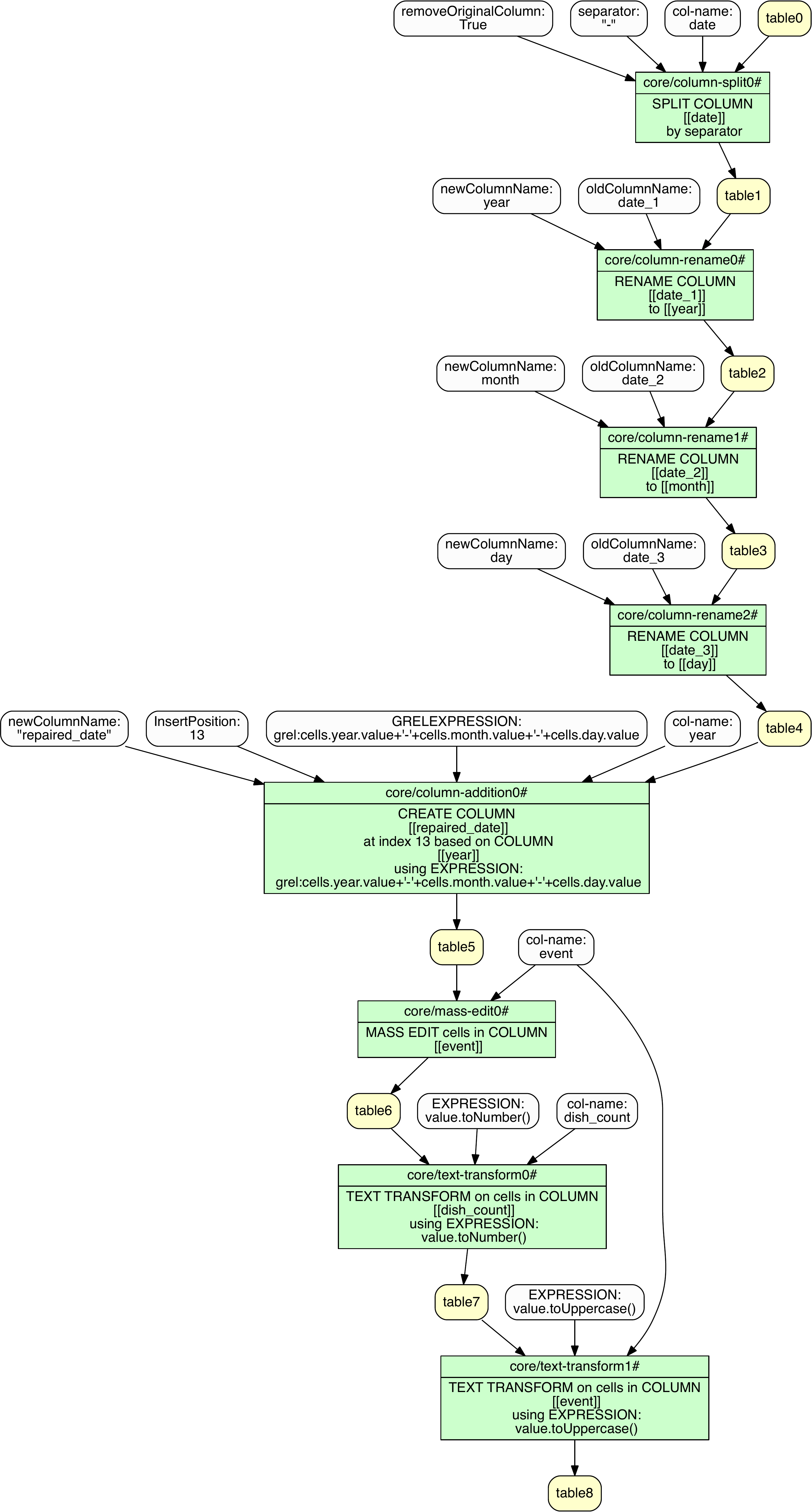}}
  \end{minipage}
  \begin{minipage}{.35\linewidth}
    {\includegraphics[height=0.32\textheight]{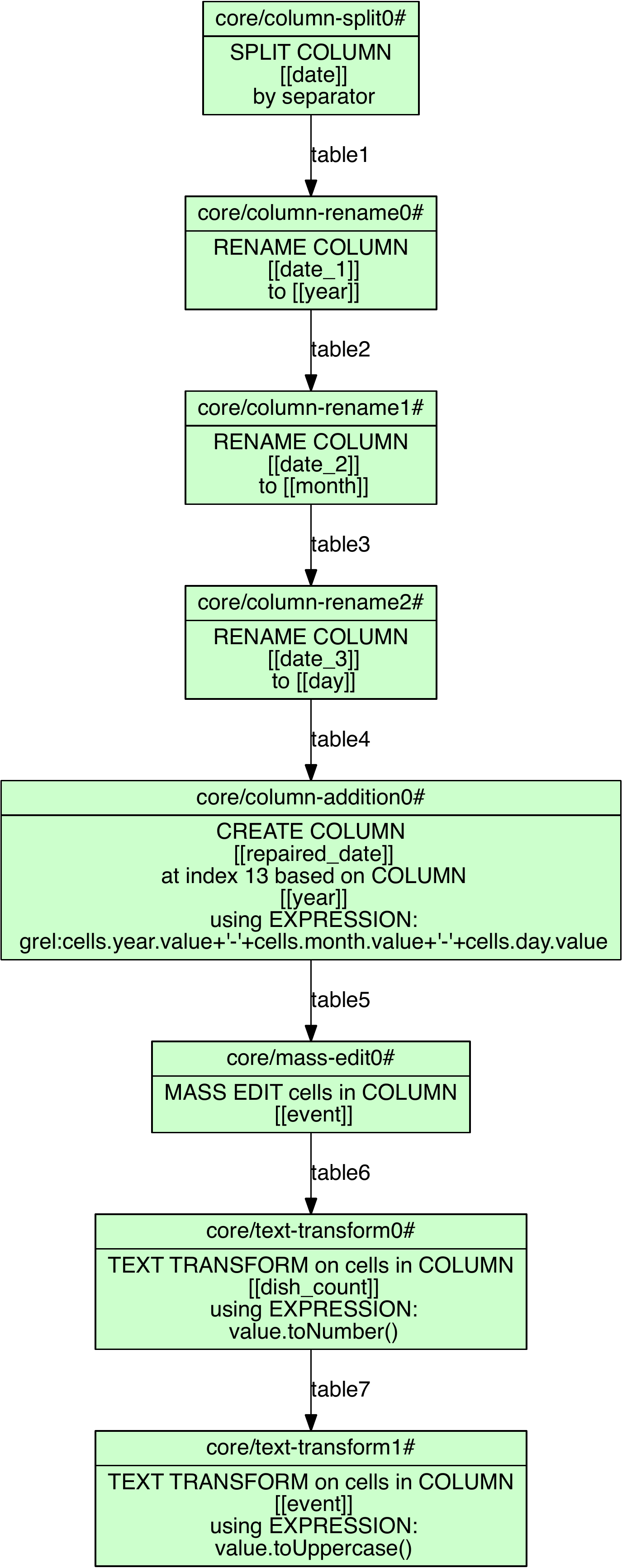}}
  \end{minipage}
  \begin{minipage}{.1\linewidth}
   {\includegraphics[height=0.34\textheight]{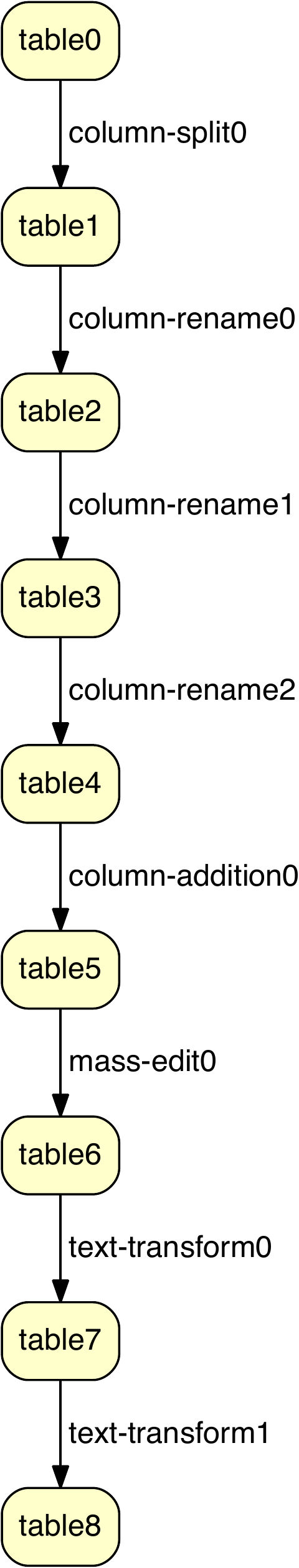}}
 \end{minipage} 
\caption{\textbf{Sequential (Linear) YesWorkflow model} of the JSON-recipe depicted in Fig.\,\ref{fig:recipe}. The \emph{combined view} (left) depicts the YW model consisting of data transformation \emph{steps} (green boxes), \emph{data} nodes (yellow), and \emph{parameter} nodes (white). By omitting certain details, the \emph{process view} (center) and \emph{data view} (right) emphasize steps and data, respectively. For the sequential model the data granularity is ``table'', i.e., each snapshot $\mathsf{table}_{n+1}$ depends on (1) the preceding snapshot $\mathsf{table}_n$, (2) a transformation operation, and (3) the transformation parameters.}
    \label{fig:linear}
\end{figure}
 
The OpenRefine operation history is linear by nature, as it reflects the user's sequential execution of interactive data cleaning steps. In the linear workflow model, we simply view this sequence of operations as a dataflow pipeline in which transformation steps $S_i$ and ``data snapshots'' $D_j$ alternate. For example, the exported recipe from Fig.\,\ref{fig:recipe} gives rise to a linear 8-step workflow: 
 \begin{displaymath}
   (D_0) \to [S_1] \to (D_1) \to \cdots \to [S_8] \to (D_8)
 \end{displaymath}

 The actual \ortoyw\ workflow output for the linear model is shown in Fig.\,\ref{fig:linear}, using a small sample of the \emph{menus dataset} \cite{whatsonmenu} that was crowd-sourced by volunteers for the New York Public Library.
 The \emph{combined view} on the left in Fig.\,\ref{fig:linear} depicts the alternation of data snapshots $\mathsf{table}_i$ with OpenRefine steps such as \textsf{column-split}, \textsf{column-rename}, etc. The \emph{process view} (in the center) suppresses data elements, thus highlighting the eight data transformation steps instead. Finally, the \emph{data view} on the right of the figure highlights the simple linear sequence of table snapshots. In this variant of a data view, edges are labeled with the operations applied to the data tables.  

Note that this simple linear model is coarse-grained in the sense that data elements of workflows in this model represent different states (``snapshots'') of the data table, as it undergoes a sequence of data transformations. As a consequence, this model does not take into account that different steps may in fact be independent from each other when considering a finer-grained data model, e.g., based on columns rather than the whole table.

\begin{figure}[t]
  \centering
  \begin{minipage}{1.0\linewidth}
{\includegraphics[width=1.0\textwidth]{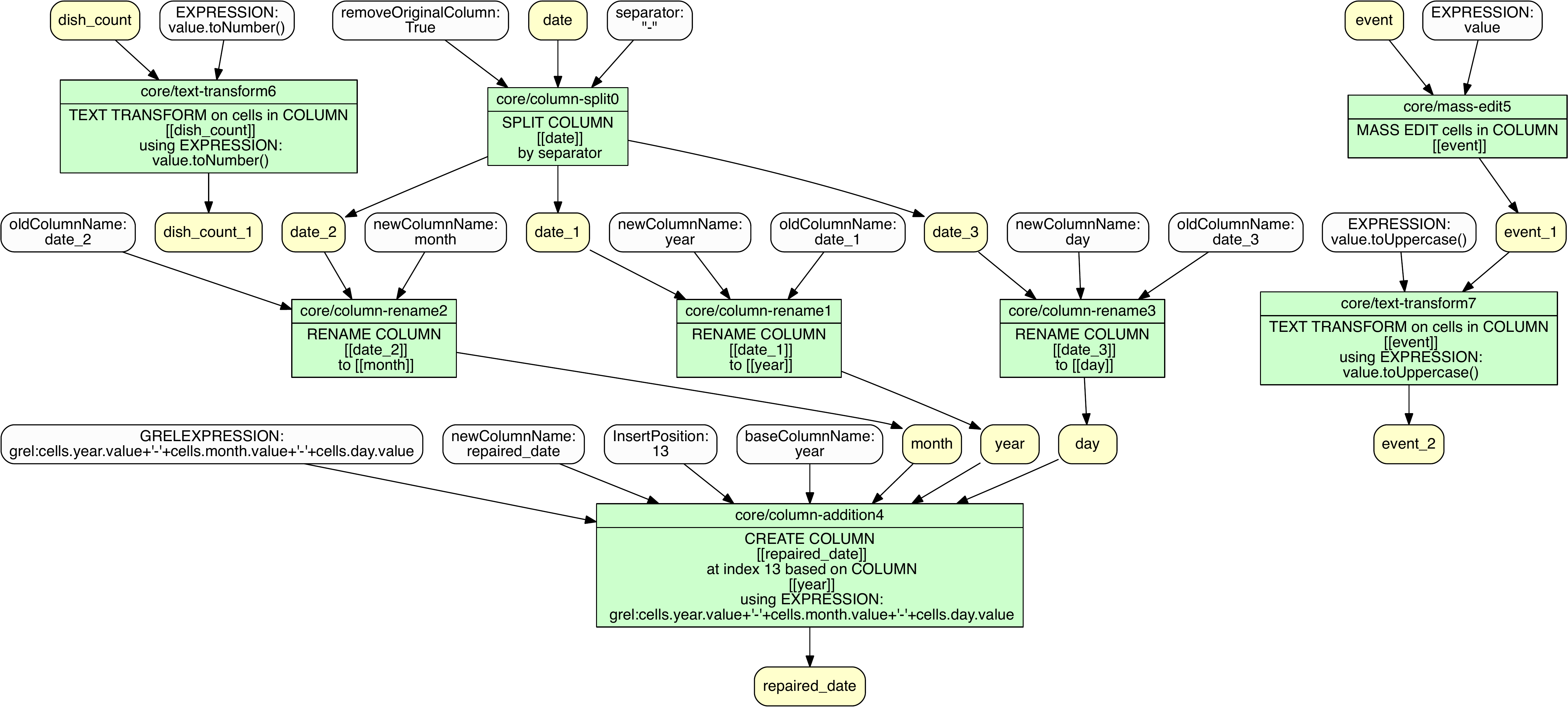}}
  \smallskip
\end{minipage}
\begin{minipage}{0.5\linewidth}
{\includegraphics[width=1.0\textwidth]{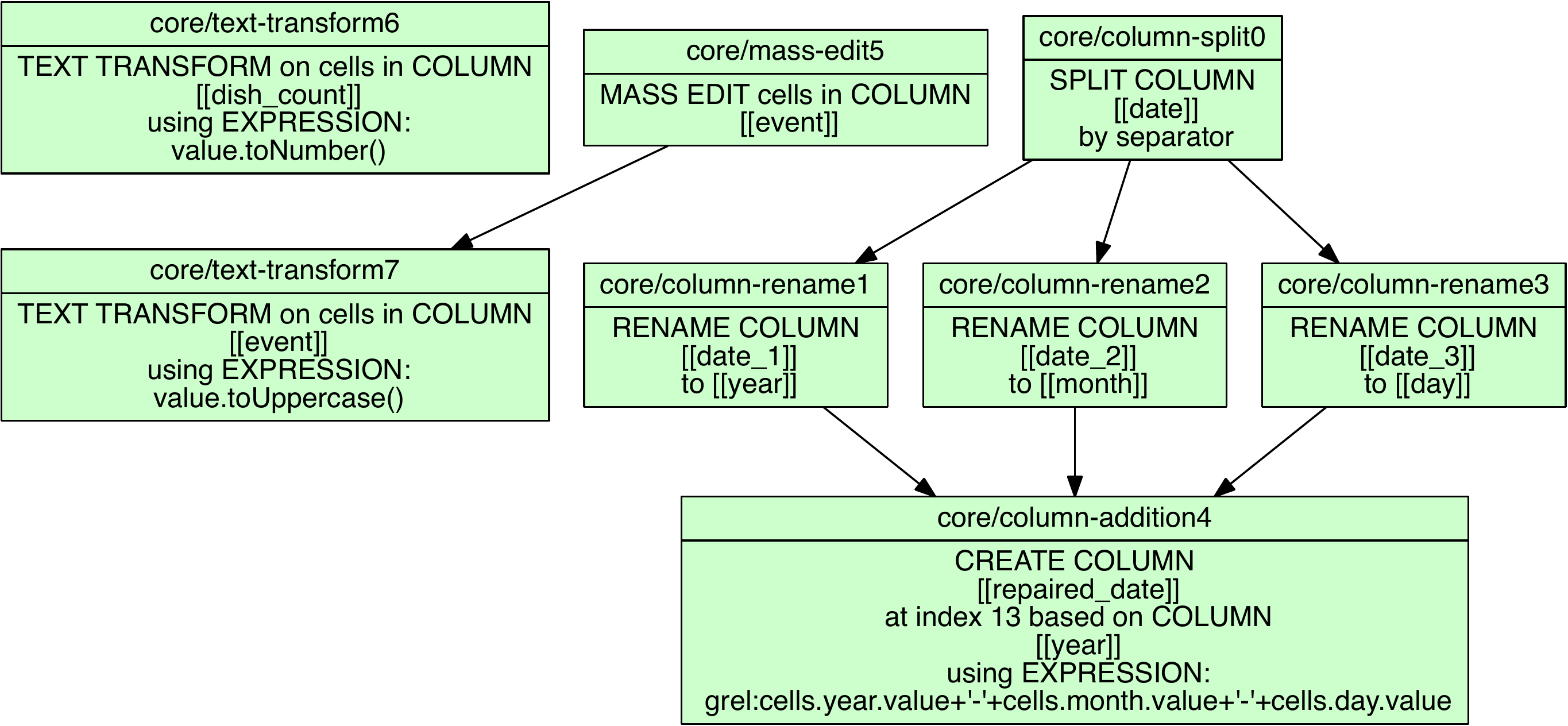}}
\end{minipage}
\hfill
\begin{minipage}{0.4\linewidth}
{\includegraphics[width=0.8\textwidth]{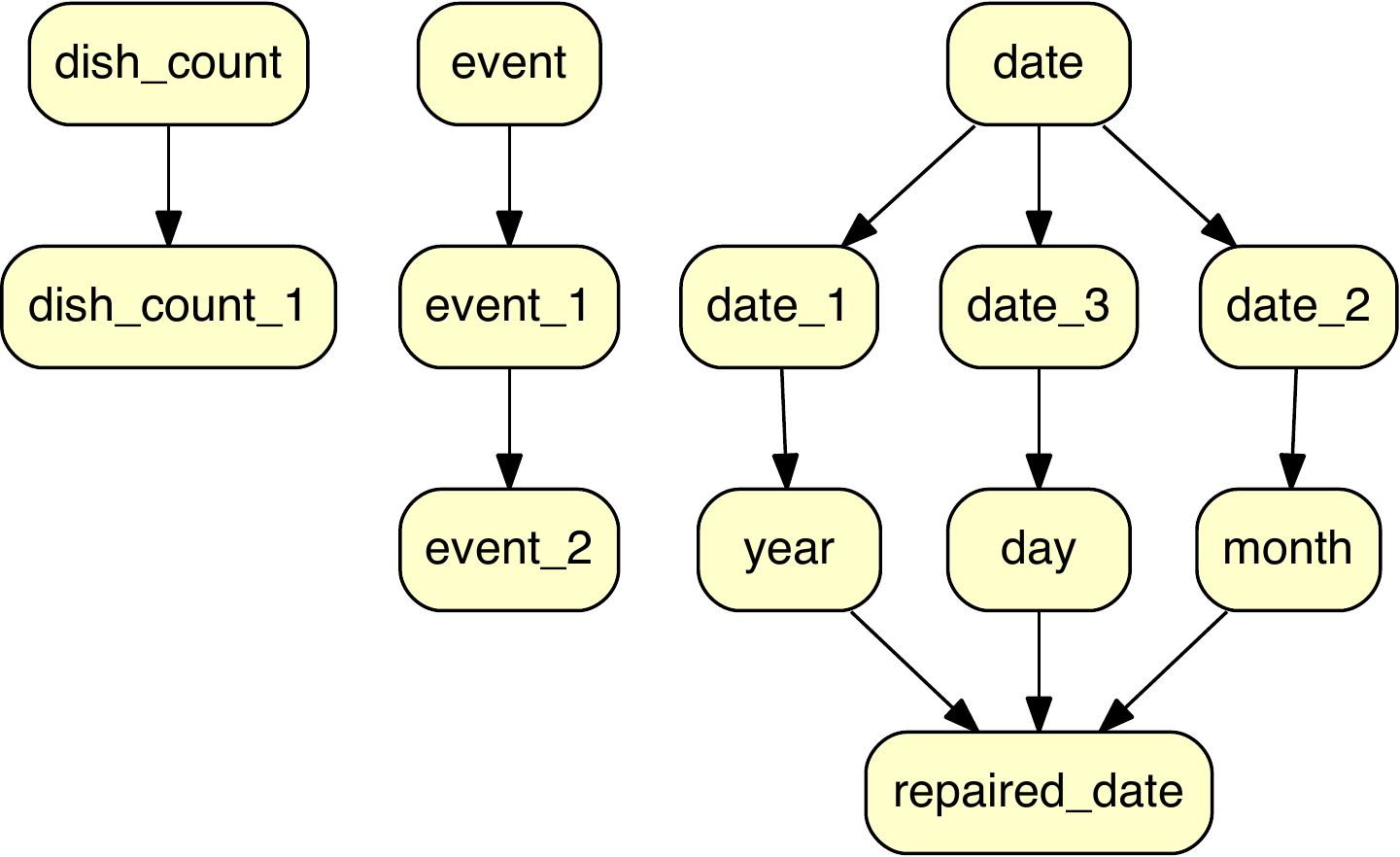}}
\end{minipage}
\caption{\textbf{Parallel YesWorkflow model} of the JSON-recipe depicted in Fig.\,\ref{fig:recipe}. The \emph{combined view} (top) depicts a detailed YW model. The {process view} (lower left) emphasizes workflow steps, while the {data view} (lower right) emphasizes data derivations. In all three YW diagrams, the independent, parallel nature of execution paths and data derivations is visible. For the parallel model the data granularity is ``column'', thus allowing a finer-grained dependency analysis relative to the sequential model in Fig.\,\ref{fig:linear} (which uses ``table'' data elements instead).}
  \label{fig:parallel}
\end{figure}

 \subsection*{The Parallel Model}

 In order to achieve a finer-grained dependency resolution between steps, the \ortoyw\ parallel model considers individual columns $C_i$ as the basic data elements. In this way, when analyzing a recipe, the tool can analyze operations at the column level (rather than at the table level) and thereby reveal data independence in a data cleaning workflow through parallel branches. To this end, \ortoyw\ employs a simple rule: if the data transformation functions $f_i$ and $f_{i+1}$ of two consecutive steps $S_i$ and $S_{i+1}$ are independent of each other, i.e., if the output columns of $S_i$ do not overlap with the input columns of $S_{i+1}$ and vice versa, then the associated functions commute, i.e.,
 \begin{displaymath}
f_i\circ f_{i+1} = f_{i+1}\circ f_i ~.    
 \end{displaymath}

\noindent When \ortoyw\ detects such a situation, it creates two new branches in the parallel workflow model, one for $S_i$ and one for $S_{i+1}$.

Fig.\,\ref{fig:parallel} depicts the parallel YW model generated by \ortoyw\ for the example from Fig.\,\ref{fig:recipe}: the combined view (top of  the figure) contains alternating sequences of the basic data elements (here: columns) and the transformation steps operating on those columns. Note that this view contains three independent subworkflows.\footnote{In this particular example, the layout algorithm of Graphviz seems to ``entangle'' the three independent subworkflows -- but a closer inspection reveals that there are indeed three independent subworkflows.} The largest of these exhibits a \emph{split-and-merge} pattern in Fig.\,\ref {fig:parallel} consisting of three branches. This pattern can be seen even more clearly in the reduced views, i.e., in the process view (lower left) and in the data view (lower right).

In addition to the ``process story'' which is evident from the process view (and the combined view), the data view tells the ``data story'' quite nicely:

There are three separate and independent data derivations, originating with the \textsf{dish\_count}, \textsf{event}, and \textsf{date} columns, respectively. The data derivation subgraph starting from \textsf{date} has three parallel branches. Inspection of the combined view (or the process view) reveals that a \textsf{split-column} operation is generating those parallel branches, one each for \textsf{day}, \textsf{month}, and \textsf{year}.  These three columns are then merged into a single new column \textsf{repaired\_date} which represents the final output of the \textsf{date} subworkflow.

Internally, the \ortoyw\ tool employs different data structures and node types to represent the various workflow patterns. These include \emph{parallel nodes} to represent independent (disconnected) subworkflows, \emph{split nodes} to represent the creation of new columns from an existing one, and \emph{merge nodes} to represent the combination of dataflow branches into a shared branch.

\subsection*{The Collapsed Model}
This is an \ortoyw\ option that can be used to collapse large groups of similar operations. Experience in a limited data cleaning study has demonstrated that for some use cases involving \textsf{clustering} and \textsf{mass-edit} operations, the operation histories can contain very long sequences of variants of essentially the same data cleaning operation. In the resulting workflow model, these long sequences (sometimes several hundred steps long) of highly repetitive steps can be more succinctly represented by collapsing highly similar operations into a new node type, representing a summary of the collapsed sequence of steps. When abstracting or collapsing a subworkflow into a single node, a separate file is produced, representing the structure of the subworkflow, so that the detailed process information can be checked if needed.



\section{Conclusions}

We have presented \ortoyw, a novel open-source tool that can automatically generate YesWorkflow annotations (and thus YW models) from OpenRefine data cleaning recipes, extracted from operation histories. When used in conjunction with YesWorkflow and Graphviz these workflow graphs can be queried and/or visualized. Here we have focused on the generation of two different kinds of YW models: a coarse-grained \emph{linear model}, and a more fine-grained, column-oriented \emph{parallel model}. The various workflow views generated by \ortoyw\ in conjunction with YesWorkflow allow users to document their data cleaning workflows in a more transparent manner, thereby also facilitating reuse of data cleaning recipes.

Numerous lines of further research and development exist. For example, in the current work we have mainly focused on generating useful workflow and prospective provenance visualizations for OpenRefine recipes. In addition, we can expand what information we extract from OpenRefine, e.g., retrospective provenance contained in OpenRefine project files \cite{li2019towards}. This would allow us to populate a hybrid provenance model, combining both prospective and retrospective provenance information. The resulting integrated model allows users to answer more complex and powerful provenance queries over their OpenRefine histories and update logs.

Another line of research is to investigate the use of alternative dataflow and data cleaning models, e.g., the formal models developed in \cite {delpeuch2020complete} and \cite{nunez2020first}.




%


%
%
\bibliographystyle{splncs04}

\bibliography{newrefs}

\end{document}